\documentclass[twocolumn,showpacs,epsfig,amssymb,preprintnumbers,floatfix,prl]{revtex4}
\usepackage{graphicx}
\usepackage{ulem}
\begin{document}\newcommand{\btau}{\mbox{\boldmath{$\tau$}}}
\newcommand{\bomega}{\mbox{\boldmath{$\omega$}}}

\title{Dynamics of a bouncing dimer}

\author{S. Dorbolo,$^{\dag,\ddag}$ D. Volfson,$^{*}$ L. Tsimring,$^{*}$ and A. Kudrolli$^{\dag}$}

\affiliation{\dag Department of Physics, Clark University, Worcester, MA 01610\\
\ddag GRASP, Physics Department, University of Li\`ege, B-4000 Li\`ege, Belgium\\
* Institute for Nonlinear Science, University of California, San Diego, La Jolla, CA 92093}

\date{\today}

\begin{abstract}
We investigate the dynamics of a dimer bouncing on a vertically
oscillated plate. The dimer, composed of two spheres rigidly connected
by a light rod, exhibits several modes depending on initial and
driving conditions. The first excited mode has a novel horizontal drift
in which one end of the dimer stays on the plate during most of the
cycle, while the other end bounces in phase with the plate. The speed
and direction of the drift depend on the aspect ratio of the dimer.
We employ event-driven simulations based on a detailed treatment of frictional
interactions between the dimer and the plate in order to elucidate the
nature of the transport mechanism in the drift mode. 
\end{abstract}
\maketitle

The shape of grains has a fundamental impact on static and dynamic properties of granular materials. Packing of granular
material is affected by the relative orientation of grains~\cite{packing}, the statistics of avalanching is  apparently
different in regular sandpiles and piles of elongated rice~\cite{SOC}, and strongly shaken rods exhibit
self-organization which leads to formation of large vortices~\cite{blair03,aranson03}. Thus the phenomenology of
asymmetric grains is very different from that of spherical beads and irregularly shaped sand.

In Ref.~\cite{volfson04}, using a combination of theory and experiments in a quasi-2D geometry (vibrated rods in an
annulus), we showed that the motion of rods is caused by the non-eccentric frictional collisions between the rods and
the vibrated plate. These collisions are translated into collective motion by multiple non-elastic collisions among the
rods during the flights between rod-plate collisions.  In this paper, we focus on an even simpler case of a single
anisotropic ``grain'' on an oscillated plate in order to study mechanisms of transformation of vibrational excitation
into horizontal transport in the absence of externally imposed asymmetry. Recent studies of thermal ratchets have
addressed the more straightforward situation where there is an imposed left/right asymmetry in the
substrate~\cite{derenyi98}.

In contrast with our previous work with thin rods, here we study the dynamics of a rigid dimer consisting of two heavy
spheres connected by a light rod. Unlike the rods case, the points of contact with the plate are easily determined even
when the dimer is horizontal at the time of collision (in this case, there are two well defined contact points).  We
also note that the oscillated dimer can serve as a natural generalization to the much studied system of a particle on a
vibrated plate which is a paradigm for period doubling and transition to chaos \cite{tufillaro92}. As we show here, in
addition to those regimes, the oscillated dimer displays spontaneous symmetry breaking and ratchet-like transport in a
system without externally imposed asymmetry.

\begin{figure}[ptb]
\begin{tabular}{ll}
\includegraphics[width=7cm]{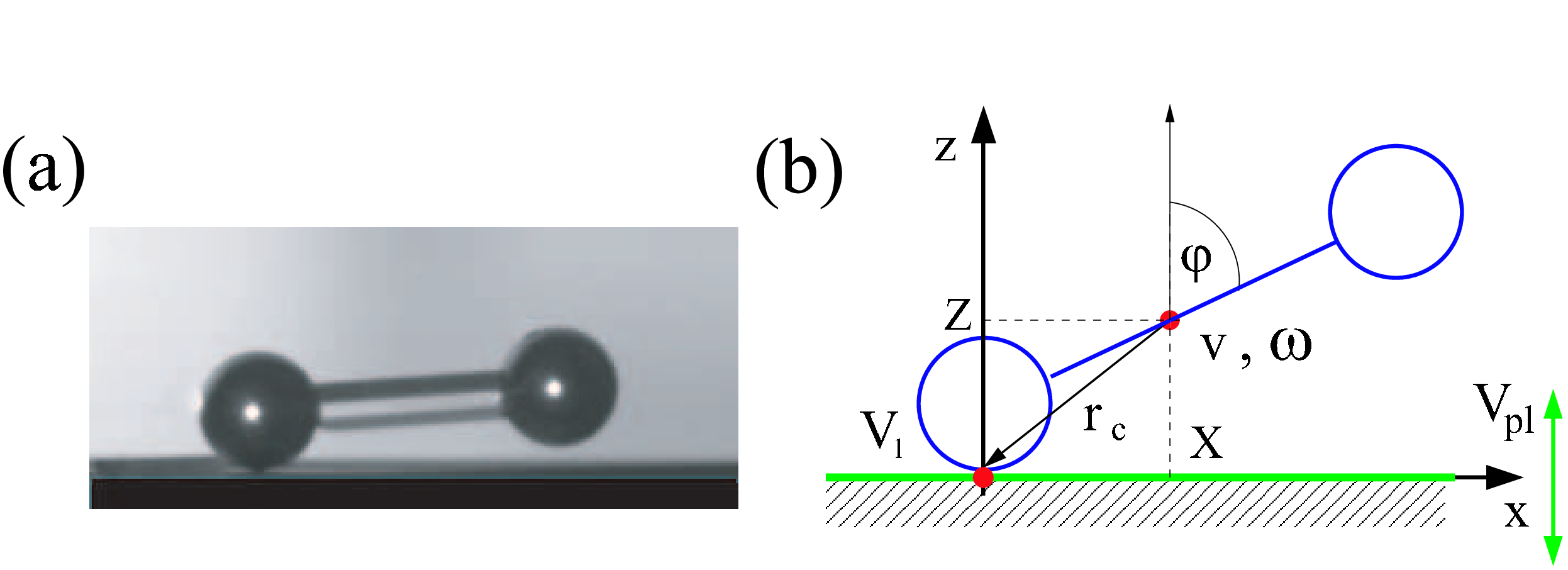} 
\end{tabular}
\caption{(a) Photo of a dimer bouncing on the vibrated plate. (b) A sketch of the system.}
\label{sys}
\end{figure}

Figure~\ref{sys} shows a photo and a schematic diagram of the system.  The experiments reported here are performed with
dimers fabricated with steel spheres with a diameter of 9.5\,mm, and a glass rod with a 4\,mm diameter. The length of
the rod is varied to change the aspect ratio $A_r$ (defined as the ratio of the length of the dimer along the symmetry
axis and the diameter of the ball)  between 2 and 10. The dimers are placed on a plate between two vertical glass walls
which are rigidly attached to the plate and separated by 10 mm in order to confine the motion to a vertical plane.  The container is vibrated with an electromagnetic shaker and a sinusoidal waveform generator. The oscillation frequency $f$ and magnitude $\Gamma = 4 \pi^2 A f^2/g$, where $A$ is amplitude of plate oscillation, and $g$ is gravitational acceleration, can be varied independently. The period of plate oscillation $T = 1/f$. The motion of the dimer and the plate is recorded using a Phantom v5.1 camera at 2000 frames per second using a standard reflection lighting configuration and a centroid algorithm which yields positions to within 20 $\mu$m.

\begin{figure}[ptb]
\begin{center}\includegraphics[width=4cm]{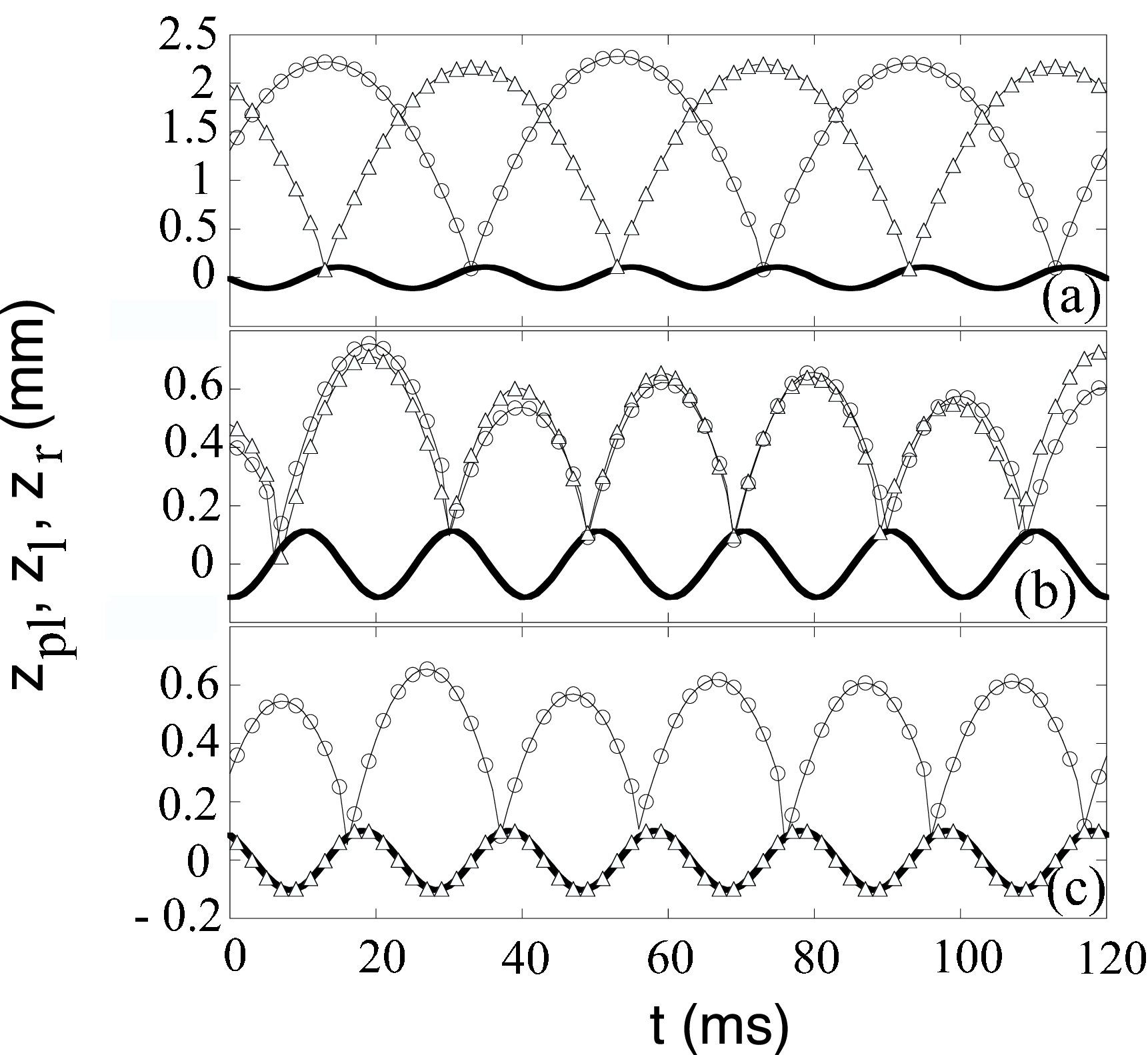}\end{center}
\caption{The vertical position of the left ($\circ$) and right ($\triangle$) tips~\cite{tips} of the dimer as a function of time for the lowest energy modes observed in the experiment. (a) The flutter (F) mode. (b) The jump (J) mode. (c) The drift (D) mode. The solid curve corresponds to the vertical position of the plate ($A_r = 3.9$, $f$ = 50~Hz, $\Gamma = 0.9$.)
}
\label{modes} \end{figure}

For $\Gamma > 1$, complicated irregular hopping and tumbling motion is observed for most initial conditions. Here, we
focus on the weak excitation regime $\Gamma <1$. Clearly, for $\Gamma < 1$ the dimer may remain on the plate.  However,
if the dimer is dropped onto the plate, three main modes of persistent motion can be observed depending on initial
conditions. The vertical coordinates of the tips of the dimer~\cite{tips} in the laboratory frame of reference are
plotted in Fig.~\ref{modes}. In the highest energy mode, the tips hit the plate out of phase at every other period of
oscillation [see Fig.~\ref{modes}(a)]. We call this motion the flutter (F) mode. In the next lower energy mode, both
tips collide with the plate once every cycle [see Fig.~\ref{modes}(b)]. We call this the jump (J) mode. In this mode
the dimer does not have significant angular momentum and resembles the first excited mode of a single particle bouncing on an oscillated plate. The lowest energy mode is shown in Fig.~\ref{modes}(c). As can be noted from the figure, one of the
ends of the dimer hits the plate once every cycle while the other appears to stay on the plate throughout the cycle. In
addition to the non-trivial vertical motion, a persistent horizontal drift is observed and therefore we call it the
drift (D) mode~\cite{movie}. Although the excitation of a particular mode depends sensitively on the initial conditions, the drift mode can be excited easily by lifting one end by a fraction of the sphere diameter and releasing it while keeping the other end on the plate. 

To test that sidewalls have negligible effect on the dynamics, we also performed experiments using a container with a shallow 1mm deep U-shaped groove with crossectional curvature which is at least four times that of the spheres in the dimer. The curvature focuses the dynamics to a vertical plane without dimer-sidewall collisions. Similar results were observed. However, higher quality position data is obtained with the container with sidewalls, and used to report details of dynamics.

We can characterize the different modes by the total mechanical energies of the vibrating dimer.  As a simple measure of such energy we take the sum of maximum elevations of both spheres above the plate. Ignoring inelasticity and horizontal translation yields the following estimates of energies per unit mass of drift, jump, and flutter modes $E_d = gT^2/8, E_j = gT^2/4, E_f = gT^2$ for $A_r\gg 1$. Therefore the ratio of the energy for the drift, jump and flutter mode is approximately 1:2:8, which is consistent with observations.

\begin{figure}[ptb]
\begin{tabular}{ll}
\includegraphics[width=7.0cm]{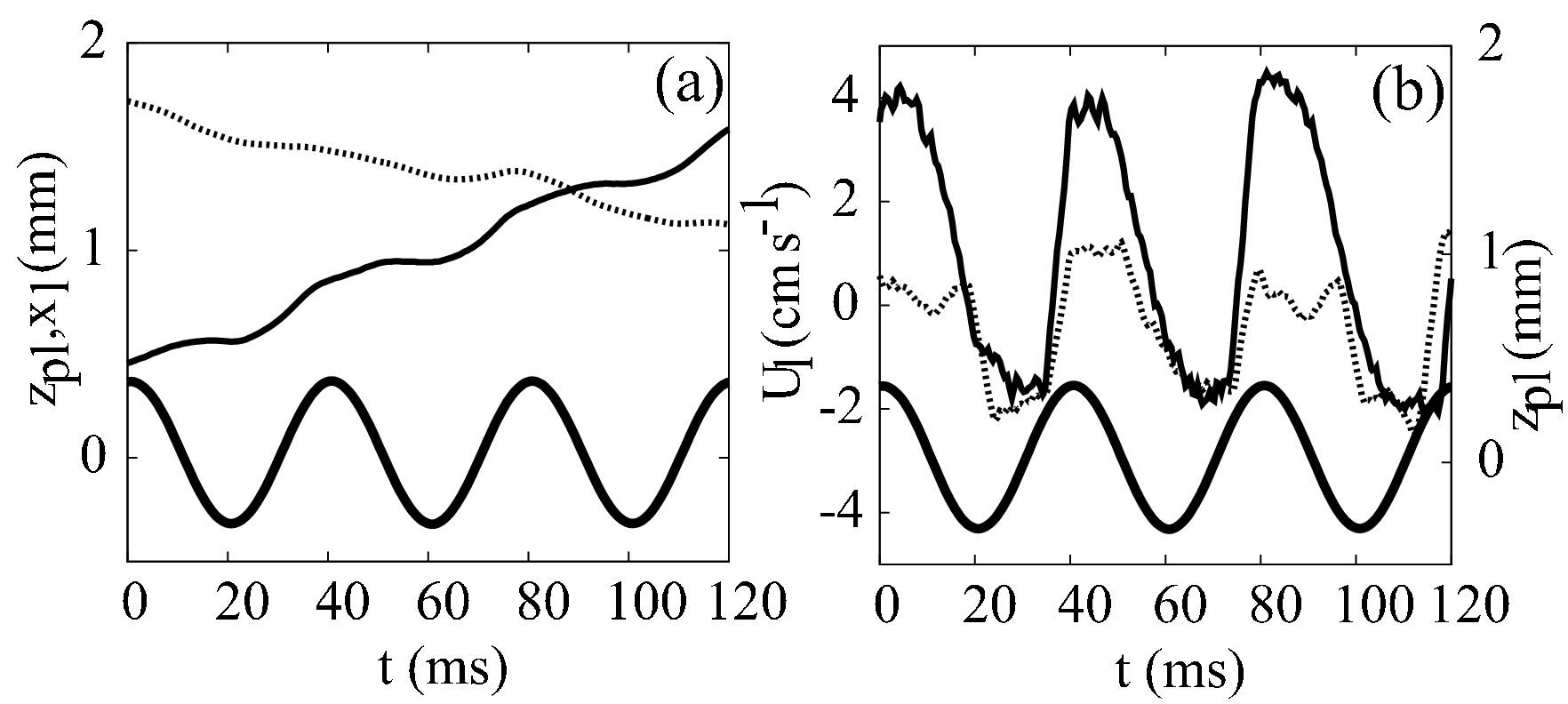}
\end{tabular}
\caption{(a) Horizontal position of the left tip of the dimer as function of time in the drift mode for $A_r=3.9$ 
(solid line) and 5.7 (dashed line), (b) The corresponding horizontal tip velocity. The thick solid line indicates 
the vertical position of the plate ($f =$ 25~Hz, $\Gamma = 0.9$). } \label{drift}
\end{figure}

In the remainder of the paper we focus on the D-mode, since it exhibits the most non-trivial dynamics. The horizontal coordinate of the left tip of the dimer ``resting'' on the plate is plotted as shown in
Fig.~\ref{drift}(a) versus time for dimer with $A_r = 3.9$ (solid line), and $A_r = 5.7$ (dashed line) at $f = 25$ Hz.
In Fig.~\ref{drift}(a), the net motion of the $A_r = 3.9$ dimer is directed towards its bouncing end. In
the following, this direction is called forward or positive.

\begin{figure}[ptb]
\includegraphics[width=6.5cm]{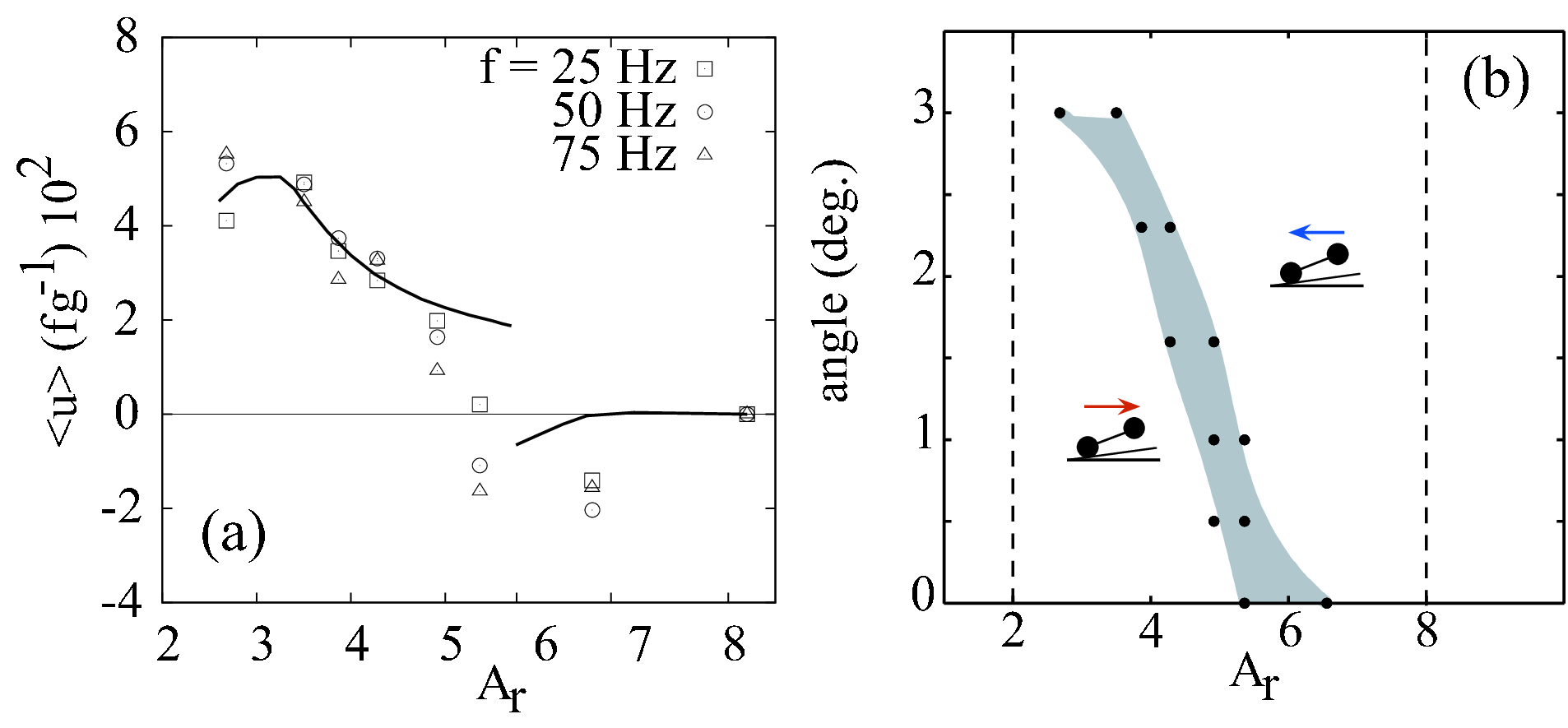}
\caption{(a) The mean horizontal speed $\langle u \rangle$ of the
dimer scaled by gravity and frequency versus $A_r$ for three values of
driving frequency. Positive direction points towards the bouncing
end of the dimer. Symbols corresponds to experiments, and the line to
the event-driven simulations at $f=25 Hz$ (lines corresponding to $f=50, 
75$ are virtually indistiguishable from this line and are not shown). Parameters of simulations are
$\Gamma=0.9$, $\epsilon=0.65$, $\mu=0.2$, $\mu_s=0.4$, and
$\hat{\mu}=0.1$, $\hat{\mu}_s=0.2$. (b) Phase diagram of the
direction of dimer translation in the D-mode as a function of
plate inclination angle and $A_r$. Robust uphill motion
is observed for a broad range of parameters. } \label{uphill}
\end{figure}

We measured the horizontal drift speed of dimers with various $A_r$ between 2 and 10 at three different frequencies of
vibration [see Fig.~\ref{uphill}(a)]. The speed is determined by measuring the time taken by the dimer to move over a
55~mm horizontal distance. For $A_r<2.5$, the D-mode becomes unstable. The drift speed is maximal for an aspect ratio of
approximately 3.5, and then observed to decrease as $A_r$ is increased. It can be also noted that dimer moves in the
backward direction at large $A_r>5$.

We also varied the driving parameters to check their influence on the dimer dynamics.  We found that the drift speed of
the D-mode is inversely proportional to $f$ [see Fig.~\ref{uphill}(a)].  However, for fixed $f$, the drift speed is almost
independent of $\Gamma$ over a broad range. For example, for $A_r = 3.9$ and $f = 25$ Hz, the measured speed was
approximately constant over a range of $\Gamma$ between 0.5 and 1.1. For smaller $\Gamma<0.5$, the D-mode spontaneously
collapses to rest, and for higher $\Gamma>1.1$, the D-mode is unstable with respect to the transition to the higher
energy J and F-modes, and their combinations. 

We checked that the drift motion is not due to any residual tilt in the plate by testing that the speed is the same when
the dimer moves in either direction. Furthermore, when the oscillated plate was deliberately tilted, {\it uphill} motion
was observed depending on $A_r$ over a range of angles [see Fig.~\ref{uphill}(b)]. (These experiments where performed with the U-shaped bottom plate which has no sidewalls.) Thus the driving mechanism in the drift mode is robust enough to overcome gravity.

Detailed examination of each tip reveals that the horizontal drift occurs due to the asymmetric slip of the low-energy
left sphere ``resting'' on the plate (see Fig.~\ref{drift}).  Specifically, the horizontal velocity $U_l$ of the left
tip is plotted in Fig.~\ref{drift}(b).  For small $A_r$, the positive slip which begins immediately after the right
sphere collides with the plate, gradually slows down and is succeeded by a slower negative slip during the phase when
the right end of the dimer approaches the plate, after which the process repeats. For large $A_r$, the positive slip is much
smaller or absent, and the drift is dominated by negative slip.

Although it appears from Fig.~\ref{modes}(c) that the left sphere is simply rolling on the plate, in fact it performs a
sequence of small jumps immediately after the right sphere collides with the plate. We confirmed both the loss of
contact after the collision during the small jumps as well as the fact that the left sphere is in contact with the plate
for a large fraction of the cycle by measuring the electric resistance between the spheres and the plate.

Our theoretical description of the dimer dynamics is based on a detailed analysis of frictional contacts between
the dimer and the plate.  Frictional inelastic oblique collision between asymmetric bodies has been
a subject of many studies in contact mechanics where various regimes of isolated contacts have been identified (see
e.g.~\cite{stronge90}). Depending on the geometry and the incident velocity of the body one can distinguish three
qualitatively different regimes: continuous slide, slip-stick, or slip-reversal (see also Ref.~\cite{volfson04} for the
case of thin rods).  In addition to these regimes, the dimer in the D-mode experiences an even greater variety of
contacts with the plate, including three types of double collisions (slide, slip-stick, and slip reversal) when two
spheres touch the plate simultaneously, and two regimes of rolling (with or without sliding).

The dynamics of the dimer is governed by the Newton
equations written  for the center of mass (CM)
velocity ${\bf v}=(u,0,v)$ and the angular velocity $\omega$
\begin{equation}
m\, \dot{\bf v} = \sum_c {\bf F}^c + m{\bf g}, \qquad
I\, \dot{\bomega} = \sum_c {\bf r}_c \times {\bf F}^c, \label{eqw}
\end{equation}
where $m, I$ are the dimer mass and moment of inertia,
${\bf F}^c=(F_x^c,0,F_z^c)$ is the contact force, and ${\bf
r}_c=(-X_c,0,-Z_c)$ is radius-vector from CM to a contact point (CP) [see also Fig.~\ref{sys}].
The sum over  $c\in \{l,r\}$ indicates that either left or right or both
spherical ends can be in contact with the plate during collisions.  The CP
velocity components in the plate frame of reference are given by
$U_c=u-\omega Z_c,\ V_c=v+\omega X_c - V_{pl}$, where $V_{pl}(t)$ is the plate
velocity.
In a sliding contact, $F_x^c=-j\mu F_z^c$, where $\mu$ is the friction
coefficient and $j=\mbox{sgn} U_c$, and in the sticking phases, $F_x^c$ is found from
the no-slip condition $U_c=0$. Transitions from stick to slip occur when
$|F^c_x|=\mu_s F^c_z$ with the static friction coefficient $\mu_s$.
Although the gravitational force can be neglected during collisions, it must
be taken into consideration during flight and rolling.

For isolated collisions, we integrate Eqs.~(\ref{eqw}) with
initial conditions $(u^b,v^b,\omega^b)$ and the kinematic
condition $V_c^a=-\epsilon V_c^b$ (where indices $b,a$ correspond
to before and after collision, and $\epsilon$ is the kinematic
restitution coefficient), to obtain after-collision CM velocities
$(u^a,v^a,\omega^a)$. Note that the dynamic and static
friction coefficients for the rolling phase in general can differ
significantly from those for the short collisions \cite{israelachvili} and
we use separate notations $\hat{\mu}$ and $\hat{\mu}_s$ for the latter coefficients.
A detailed description of all contact types is rather
cumbersome and will be presented elsewhere~\cite{dorbolo04b} (see also
Supplementary material).

\begin{figure} \begin{center}
\includegraphics[width=6cm]{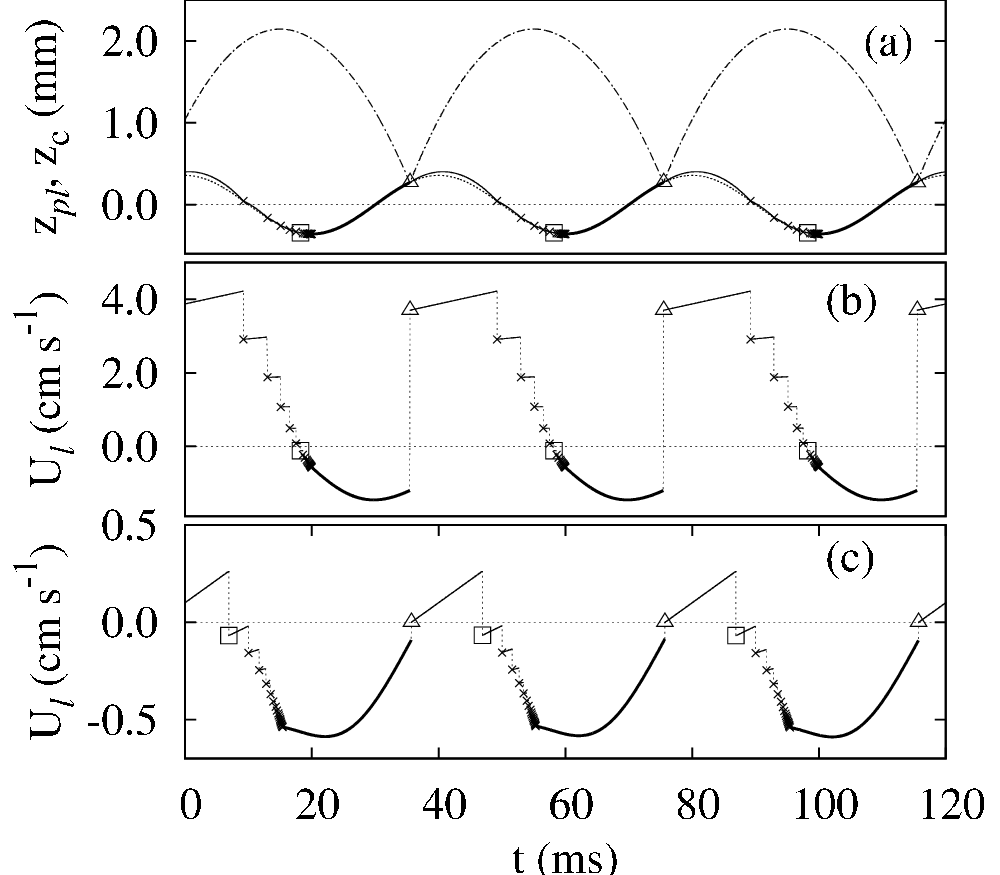}
\caption{
Vertical coordinates of left and right tips of the dimer (a)
and the horizontal velocity of the left contact point for the short dimer
$A_r=3.9$ (b) and long dimer $A_r=6.1$ (c). Different
symbols and line styles denote different modes of motion and collisions
between the dimer and the vibrating plate:
single slide ($\times$), single slip reversal ($\square$), double
collision ($\triangle$), rolling (thick solid line), flight (thin solid
line). Parameters of simulations are the same as in Figure \ref{uphill}. }.
\label{fig:theory}
\end{center}
\end{figure}

We combined the conditions for the occurrence of the various kinds of collisions and the forces acting on the dimer
into an event-driven algorithm. We used this algorithm to elucidate the origin of the observed regimes of dimer motion.
In D-mode, the left tip rolls on the plate with negative slippage as the right tip approaches
the plate (see Fig.~\ref{fig:theory}).  At time $t_0$, the right tip collides with the plate with
an incident contact velocity $(U^b_r<0, V^b_r<0)$.  In the first phase of collision, negative
slippage slows down due to the frictional force $F_x^r=\hat{\mu} F_z^r$. For large
enough friction, at a
certain time $t_0'>t_0$ during collision, the negative slip stops
($U_r(t_0')=0, V_r(t_0')=V_r^s$). If formally $V_r^s>-\epsilon V^b_r$, the
dimer leaves the plate before the negative slip stops, and so the
collision would be of a continuous slide type. Continuous negative slide cannot persist in
the stationary regime, because this would yield cumulative positive horizontal impulse
transfer from the plate to the dimer over the period of vibration and hence an accelerated motion.
For $V_r^s<-\epsilon V^b_r$, there can be two possible scenarios for the second phase of the double collision.
For large enough aspect ratios, $\hat{\mu}_s(A^2+7/5)>A$, (where  $A=A_r-1$,) the CP sticks at $t>t_0'$, i.e.
$U_r(t>t_0')=0$. For $\hat{\mu}_s(A^2+7/5)<A$, the slip reverses direction, in which case it can be
shown that $U^a_r$ is given by
\begin{displaymath}
\frac {5\hat{\mu} A^2-5A+7\hat{\mu}}{10 A^2-5\hat{\mu} A+2}\left[(1+\epsilon) V_r^b
-\frac {10 A^2+5\hat{\mu} A+2}{5\hat{\mu} A^2+5A+7\hat{\mu}}U_r^b\right].
\end{displaymath}

For small $A_r$, $U^a_r$ becomes positive after the double collision in agreement with experiment
[cf. Fig.~\ref{fig:theory}(b) and Fig. \ref{drift}(b)]. This ``kick'' during
the two point collision is the main source of the positive drift momentum.  Simultaneously, the right tip acquires a
finite vertical velocity $V_r^a$ which equals $-\epsilon V_r^b$.  The left tip also acquires a small
positive vertical velocity $V_l^a \ll V_r^a$, resulting in the dimer being lifted off the plate.  Following the
lift-off, the left end makes a sequence of collisions with the plate which result in a decaying sequence of small flights.
This sequence converges through inelastic collapse at time $t_1$, after which the left tip of the dimer stays on the
plate. Meanwhile the right end of the dimer is still in the air, and the rolling mode with slippage commences and
continues until the right tip finally collides with the plate, after which the process repeats.

For large $A_r$, the double collision is of the slip-stick type, so that $U_c^a=0$, and the mean slip velocity is
negative [Fig.~\ref{fig:theory}(c)]. While this picture qualitatively agrees with the experimental findings [compare
Fig.~\ref{fig:theory}(b),(c) with Fig.~\ref{drift}(b)], in order to match the mean drift velocity quantitatively [see
Fig.~\ref{uphill}(a)], we had to choose $\hat{\mu}, \hat{\mu}_s$ for short-term collisions which were half of those for
the rolling phase.  While we do not have a good explanation for the smaller values, it is well
known~\cite{israelachvili} that the friction properties depend strongly on the age of the contact and therefore can
differ between short-time collisions and relatively long rolling contacts. As seen in Fig.~\ref{uphill}(a), even with
this fitting of friction parameters, there is a significant deviation between theory and observations for large aspect
ratios. In particular, experiments do not show a discontinuous transition from forward to reverse drift at a certain
$A_r$. We believe that this discrepancy can be explained by the sensitivity of the double collision to small variations
in frictional properties of the plate at large $A_r$.

In conclusion, we investigated the dynamics of a dimer bouncing on an
oscillated plate with the special focus on the first excited D-mode in which
the dimer travels in one of
the two directions selected via a symmetry breaking. In typical
fluctuation-driven transport systems, the direction of transport is
selected by the local asymmetry of microscopic potential
~\cite{derenyi98}. In contrast, our bouncing dimer presents a
novel phenomenon of ratchet motion in a system without an imposed
asymmetric potential~\cite{olafsen}.
Our theoretical analysis reveals that the D-mode
is in fact a concatenation of multiple collisions which include
individual, double, and rolling contacts. Horizontal
transport is forced by the Coulomb friction during collisions and
rolling phases.  
Because our preliminary experiments show that the drift mode survives
collisions, one can anticipate that a collection of dimers, while initially
exhibting random motion, will align mutually due to excluded volume interactions
amplified by inelastic collisions. We expect that this alignment would give rise 
to a global convection similar to vortices observed with rods~\cite{blair03,aranson03}.

SD thanks FNRS for financial support. This work was
supported by NSF grant \# DMR-9983659 (Clark) and
the U.S. DOE grant DE-FG02-04ER46135 (UCSD).


\begin{thebibliography}{99}
\bibitem{tips} We call the tip the point on the sphere surface directly underneath its center.
\end{thebibliography}
\end{document}